\documentclass[]{spie}  %>>> use for US letter paper

\usepackage{amsmath,amsfonts,amssymb}
\usepackage{graphicx}
\usepackage[colorlinks=true, allcolors=blue]{hyperref}
\usepackage{wrapfig}
\usepackage{lscape}
\usepackage{rotating}
\usepackage{epstopdf}
\usepackage{subcaption}
\usepackage{setspace}
\usepackage{lineno}
\usepackage{multirow}
\usepackage{float}
\title{Stages of commissioning alignment for three-mirror anastigmat (TMA) telescopes}

\author[a,b]{Solvay Blomquist}
\author[a,b]{Heejoo Choi}
\author[a,b]{Hyukmo Kang}
\author[b]{Hayden Kim}
\author[b]{Kevin Derby}
\author[b]{Pierre Nicolas}
\author[b]{Joanna Rosenbluth}
\author[a]{Patrick Ingraham}
\author[a]{Ewan S. Douglas}
\author[a,b]{Daewook Kim}
\affil[a]{Steward Observatory - The University of Arizona \linebreak 933 N Cherry Ave, Tucson, AZ 85721 \linebreak}
\affil[b]{Wyant College of Optical Sciences - The University of Arizona \linebreak 1630 E University Blvd, Tucson, AZ 85721}

\authorinfo{Further author information: (Send correspondence to H.C. and S.B.)\\H.C. E-mail: hchoi@optics.arizona.edu\\ S.B. E-mail: sablomquist@arizona.edu}

\begin{document} 
\maketitle

\begin{abstract}
Reliable, autonomously, deployment of telescopes enables a wide range of possible science cases.
In this paper, we present a method for multi-stage telescope alignment with a simple commercial imaging sensor. For these studies, we use a design of an example three mirror anastigmat (TMA) telescope and consider how the average spot size across the detector changes as a function of primary (M1) and secondary (M2) mirror positioning. This multi-stage alignment procedure will consist of three subprocesses, starting with a coarse alignment and converging down to a finer alignment before moving on to a stage where the telescope will refine its misalignments for data acquisition. This alignment strategy has been tested and meets ``diffraction limited'' requirements on a subset of misalignment cases from a statistical Monte-Carlo simulation given misalignment tolerances on the telescope. 

\end{abstract}


\section{INTRODUCTION}
For all telescopes, both ground- and space-based, it is important to consider how the degradation of optical performance from assembly to deployment can be compensated during the initial stages of the telescope's operation. Especially in austere environments, such as Antartica, the moon, and deep space, locations that inherently have limited accessibility for human intervention. Any potential commissioning algorithm should be hypothesized and tested rigorously through simulation and lab work ahead of the launch. All telescope alignment until the 1980s were considered passive, meaning that after initial set-up of the optical system any sort of alignment had to be done manually, until active alignment was introduced and is used for most modern day telescope alignment \cite{NOETHE20021}.  Active alignment employs a closed loop correction based on how the information from the image plane changes in response to moving different components of the telescope \cite{NOETHE20021}. From this concept of active alignment, we will use highly aberrated star images captured at the image plane of a misaligned telescope to evaluate how it evolves as a function of moving the dominant optics (primary and secondary mirrors). In this process, various strategies can be utilized to evaluate and guide the alignment procedure given some measurement being done in parallel. Ultimately, the procedure described in this paper will take a telescope from a highly misaligned state where starlight reaches the detector to diffraction limited performance, in a timely and low-cost manner. One of the main goals of this work is to show that this algorithm can correct a subset of possible misalignment states on orbit to the point at which fine alignment can take over. 
 
Often, telescope alignment has been done by analyzing images captured by science instruments for the same reasons of low cost and time efficiency \cite{nissly_efficient_2024, grey_control_1986, knight_observatory_2012, 10.1117/12.736059, acton_multi-field_2012, zhang_field-balancing_2024}. For such as the Giant Magellan Telescope (GMT), during the commissioning phase there have been simulations conducted that show effectiveness of using a combination of on-axis wavefront sensor and model-based techniques to align the telescope system \cite{mcleod2014giant, bouchez2018overview}. However, with large segmented telescopes, like the GMT and James Webb Space Telescope (JWST), segment identification is a key part of commissioning while the telescope alignment presented in these studies has a monolithic mirror. While eliminating segment identification and further intricacies of using a large segmented mirror, we will use aberrated star images from the detector plane to drive positioning of the primary and secondary mirror, a recommended design choice to be able to increase the science-to-cost ratio of the mission and also cut down the time needed for commissioning during the early stages of the mission\cite{feinberg2012space, douglas2023approaches}. The goal of this commissioning algorithm is to also be able to perform the alignment completely autonomously with minimal computing resources, letting the telescope converge given some informed model in response to how the star image behaves.   

For algorithm development and testing, we are using the TMA design outlined in Kim et al. 2023. \cite{kim2023compact} By utilizing the telescopes' wide field of view, we plan on using wide-field stars available for the alignment. With a higher sampling of field points, our alignment can be driven with more accuracy. Using the TMA design, we decomposed the available rigid body motion into three bodies - primary mirror (M1), secondary mirror (M2), and all others (tertiary mirror (M3) through image plane). In this study, we assume that the optics from M3 to the focal plane are fixed relative to the primary and secondary and want to eliminate the possibility of introducing further aberration to the system by unnecessarily moving more optics. 

\begin{figure}[hbt!]
    \centering
    \includegraphics[width=1\textwidth]{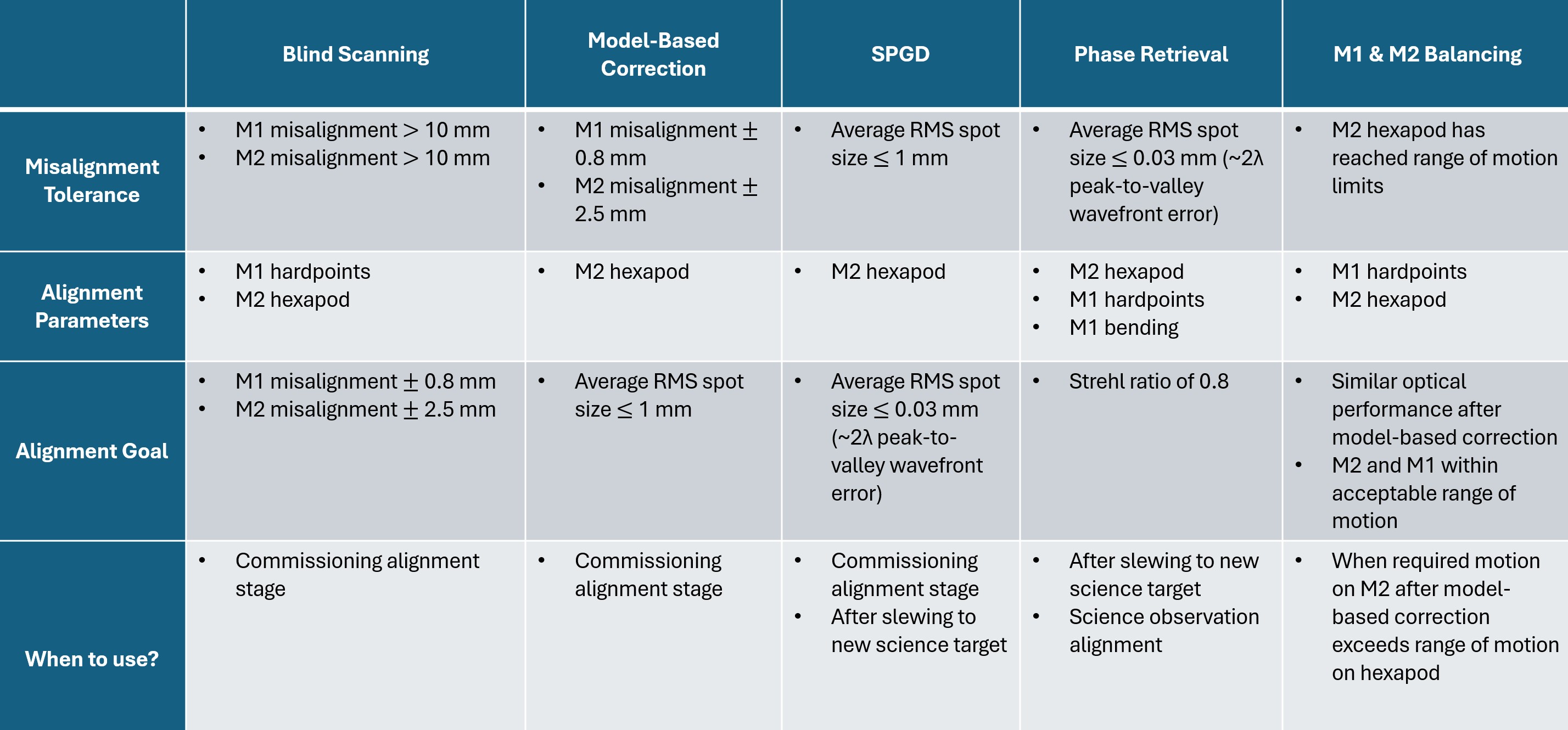}
    \caption{Summary of the commissioning steps for the telescope along with their requirements and alignment goals. For simulating the telescope alignment at each stage of the algorithm, there is a misalignment tolerance defined, alignment parameters defined for what should be adjusted at each stage, the goal state at the end of alignment and when to use the alignment step.}
    \label{commissioning_chart}
\end{figure}

The commissioning alignment further described in this paper will happen sequentially in four stages, with each stage having its own prescribed performance cut-off before moving on to the next stage. The details of each step in the commissioning algorithm are shown in Figure \ref{commissioning_chart}. The fifth column of the chart describes the M1 and M2 balancing alignment, which will only take place if the M2 correction defined by the previous stages exceeds its range of motion.  First, the telescope will undergo a Blind Scanning procedure that consists of moving M1 and M2 until the maximum amount of flux from the star field is seen on the detector. Blind Scanning will be further discussed by Kim et al. 2025 \cite{hayd_talk}. Model-based correction, SPGD and M1 and M2 balancing will be discussed in this paper and are part of the coarse commissioning alignment, fine alignment procedure will consist of phase retrieval which will be further discussed by Derby et al. 2025 \cite{kev_talk}.

For the studies presented in this paper, we will focus on testing the series of coarse alignment algorithms assuming the telescope has been corrected through Blind Scanning. Following Blind Scanning, M2 will undergo a model-based correction algorithm that consists of adjusting M2 in three translational degrees of freedom until a local minima is found from an ideal combination of M2 position in translational degrees of freedom. If M2 reaches the limits of its dynamic range of motion at this step, then a M1 and M2 balancing alignment will take place. The goal with this step is to use M1 and re-adjust M2 to move it away from its range limits. The final stage of commissioning is Stochastic Parallel Gradient Descent (SPGD), which will be used to further clean up misalignment by stochastically adjusting the available degrees of freedom on M2 (and may re-engage M1 as M1/M2 balancing if necessary) until the goal final state of the telescope is met before finer alignment. \cite{kev_talk} 

Once the spot size has converged to the target value for further alignment to take over, this marks the end of the commissioning alignment at this stage of the study. Further improvement can theoretically be made to the wavefront but there are other algorithms better suited at this stage of fine alignment, such as phase retrieval. In the fine alignment stage, the wavefront sensor directly corrects the telescope and retains image quality over the allotted exposure time defined by mission/science requirements. \cite{derby2023integrated, kev_talk}.

The measurement of spot size as a function of M2 motion will come at no extra cost to the telescope as long as an image sensor is available during the commissioning stage. This sequence of alignment processes has been tested against a subset from a Monte Carlo simulation generated from possible misalignment ranges of the primary and secondary mirror \cite{choi2023approaches}. From selected misalignment cases from the Monte Carlo analysis, we were able to successfully show that the presented algorithm converges to our defined performance cutoff in orbit.

For this study, our goal is to be able to show the effectiveness of a series of coarse alignment algorithms to take a presented TMA telescope design from a highly misaligned state post deployment to near diffraction limited performance where finer alignment could take place. In this paper, we present the methodology behind a series of alignments following the initial Blind Scanning phase where we assume a star image has been detected for use in alignment. Future works will consist of testing this algorithm on other TMA designs to conclude that this alignment can be used as a general solution once the telescope reaches its position on orbit. 

\section{Methods of Alignment} 

The primary and secondary mirrors of the TMA both have six degrees of freedom that can be utilized to correct the proposed telescope in orbit. The six degrees of freedom are translation (or decenter), ($D_x, D_y$ and $D_z$) and tip, tilt or clocking ($T_x, T_y$ and $T_z$). Please, note that, depending on the surface type of the mirrors (e.g., off-axis conic surface), the theoretical degrees of freedom of a mirror could be less than six. We will not take $T_z$ into account for this study, as that includes clocking the primary or secondary mirror and with the current optical design, there is no asymmetry about the Z-axis (i.e., optical axis) so we do not need to consider changing $T_z$ as a function of average spot size. In this section, we discuss the methodology behind the stages of alignment in the order in which they take place after deployment. These alignments are performed under the assumption that light is reaching the image plane from a distant star and we can roughly detect a blob star image following Blind Scanning. The main metric that will be used when assessing alignment throughout these alignment phases will be the averaged root mean square (RMS) spot size across the detector. Each stage of alignment throughout the telescope's commissioning phase will work to converge the RMS spot size to a defined value. Once the threshold defined in Figure \ref{commissioning_chart} is met, the primary and secondary mirror positions are recorded and passed on as an input for further alignment needed for science observations including phase retrieval and bending of the primary mirror. \cite{derby2023integrated, blomquist2023analysis, kev_talk}

The first step of the alignment is Blind Scanning, consisting of moving M1 and M2 until the maximum total flux from the star field is captured at the image plane\cite{hayd_talk}. The assumption of this step is that the size of the star image will indicate whether or not the telescope is approaching collimation. At this stage, traditional wavefront-based techniques often fail, especially in cases of severe mirror misalignment. Using the statistics from the star field images around the telescope, we will drive the primary and secondary mirror (M1 and M2) alignment with big steps until the maximum amount of star intensity is seen on the detector. Further discussion of this step in commissioning will be discussed in Kim et al. 2025 \cite{hayd_talk}.

From there, we can begin coarse alignment by moving M2 in the three translational degrees of freedom as a function of the spot size across the detector until the average RMS spot size across the image plane is below $1$ mm. This current cut-off value is chosen in a way that utilizes translation to find the local minima of the system while preserving tip and tilt adjustability. The model-based correction presented will evaluate the average RMS spot size across the detector and scan M2 along the available translational axes until a minimum is found. This minima represents the tentative optimal position under the recorded M2 position in its available degrees of freedom. After the model-based correction stage, if the reported M2 position that corresponds to below $1$ mm average RMS spot size exceeds the mirror's range of motion, we will engage M1 to balance the optical performance and alleviate the range of motion on M2. Moving M1 consists of only rigid body movement in the five degrees of freedom available to optically balance out the misalignment of M2. One of the benefits of using a monolithic primary mirror is that we will not need to consider relative segment positioning and co-phasing after moving M2 which has been a critical step in prior commissioning algorithms for large meter class telescopes. After the model-based correction step, and the potential balancing of M1 and M2, the following alignment will then be initiated involving SPGD. During SPGD correction, we will randomly perturb the five degrees of freedom on M2 while measuring average spot size across the detector. The direction and magnitude of these corrections will be determined by the trend of the spot size and the degree of freedom being aligned. 

When developing the three stages of the commissioning alignment following Blind Scanning, we invoked various tests to run each algorithm independently. Following these tests, we then put all stages together to test as one cohesive commissioning flow. 

\subsection{Model Based Alignment} \label{model_based}

 For our simulation case, we start with some misalignment case defined by the ranges defined in Table \ref{tab:misalign_range}. In Ansys Zemax OpticsStudio 2024 (Zemax), we can apply these ranges to the tolerance data editor and create a Monte Carlo distribution of misalignment cases. These ranges have been previously calculated and are listed in the tolerance and error budget developed specifically for this telescope design\cite{choi2023approaches}. 

 \begin{table}[h!]
    \setlength{\tabcolsep}{10pt} % Default value: 6pt
    \renewcommand{\arraystretch}{1.5} % Default value: 1
    \centering
    \begin{tabular}{|c|c|c|} \hline 
         \textbf{Degree of Freedom (DoF)}& \textbf{M2 Misalignment (units)} & \textbf{M1 Misalignment (units)}\\ \hline 
         Decenter X & $\pm$ 2.5 (mm) & $\pm$ 0.8 (mm) \\ \hline 
         Decenter Y & $\pm$ 2.5 (mm) & $\pm$ 0.8 (mm) \\ \hline 
         Decenter Z & $\pm$ 2.5 (mm) & $\pm$ 0.8 (mm) \\ \hline 
         Tilt X & $\pm$ 0.4 (deg.) & $\pm$ 0.01 (deg.) \\ \hline 
         Tilt Y & $\pm$ 0.4 (deg.) & $\pm$ 0.01 (deg.) \\ \hline 
    \end{tabular}
    \caption{Notional misalignment ranges applied to M1 and M2 before beginning the model-based algorithm. These misalignment ranges will be the goal of the Blind Scanning process before beginning the model-based fitting and correction. These values are derived from the tolerance and error budget for this TMA design \cite{choi2023approaches}.}
    \label{tab:misalign_range}
\end{table}

The first degree of freedom we will be focused on correcting is Decenter Z (separation between M1 and M2 along optical axis). The minimum for this iteration of the algorithm is found by evaluating the trend of the RMS spot size as a function of the position of the M2 Decenter Z in 10 step increments. As shown in Eq. 1, if the trend shows a clear trough within the 10 step trial, the best available Z position values can be calculated via the relationship $R(z)$ as a function of M2 position in Decenter Z, $z$. From there, the best position of M2 in Decenter Z is calculated via the vertex, as shown in Eq. 2. If the combination of M2 misalignment doesn't reach a clear trough, the trend is linear. If a linear trend is detected, we keep the minimum value at the first 10 step trial and revisit the alignment later once the combination of the misalignment is mitigated. The goal of this step of correction is not to have a perfect alignment but to get close enough for SPGD to effectively clean up remaining errors. 

\begin{equation}
    R(z) = az^2 + bz + c \\
\end{equation}

\begin{equation}
    D_z = \dfrac{-b}{2a}
\end{equation}

The Decenter Z alignment is then followed by Decenter Y and Decenter X. The same algorithm for determining the minima for Decenter Z is used for both Decenter Y and Decenter X. When calculating the best position in $D_x$ and $D_y$, Eqs. (1) and (2) are used again but replacing the variables with $x$ and $y$ to calculate $R(x)$ and $R(y)$, respectively. It should be noted that there is asymmetry in the telescope design about the Y-axis so the Decenter Y misalignment is much more sensitive compared to the Decenter X misalignment. If at any point throughout the algorithm the threshold of less than 1 mm RMS spot size is met, then this stage of the algorithm is complete and the position of M2 is recorded for SPGD.

\subsection{Primary and Secondary Mirror Balancing}

Until this point in the alignment, we utilize M2 hexapod motion while keeping the primary mirror stationary. However, this method of alignment will encounter the limit of M2 adjustability since M1's misalignment is also dominant source of the degradation in optical performance. After the required motion of the secondary mirror is calculated, we then evaluate how much the mirror has moved in relation to its range of motion. If M2 reaches the limits of its range of motion, we will engage M1 to balance out M2 and alleviate the strain on the hexapod's motion while maintaining the optical performance of the telescope at the end of model-based correction.

\begin{figure}[hbt!]
    \centering
    % Top row: Two subfigures
    \begin{subfigure}{\textwidth}
        \centering
        \includegraphics[width=1\textwidth]{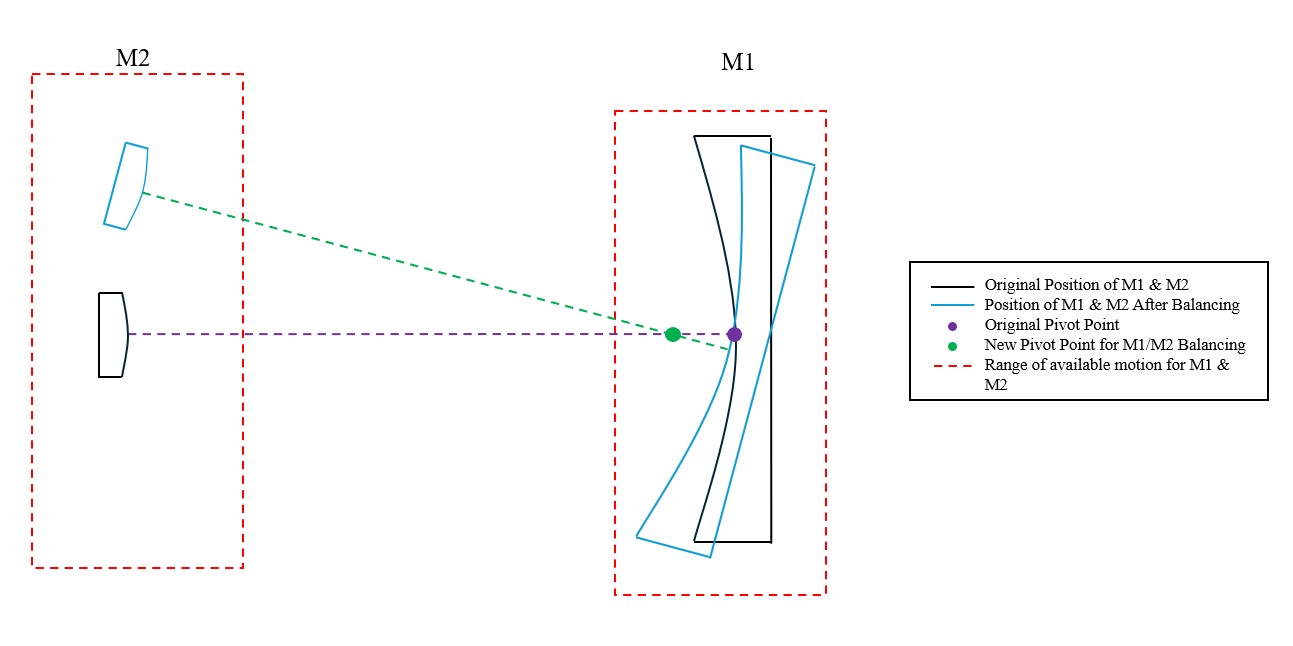}
        \caption{Pivot point orientation between M1 and M2 in the case where M1 has a large range of motion to be able to balance out M2's misalignment.}
    \end{subfigure}
    \begin{subfigure}{\textwidth}
        \centering
        \includegraphics[width=1\textwidth]{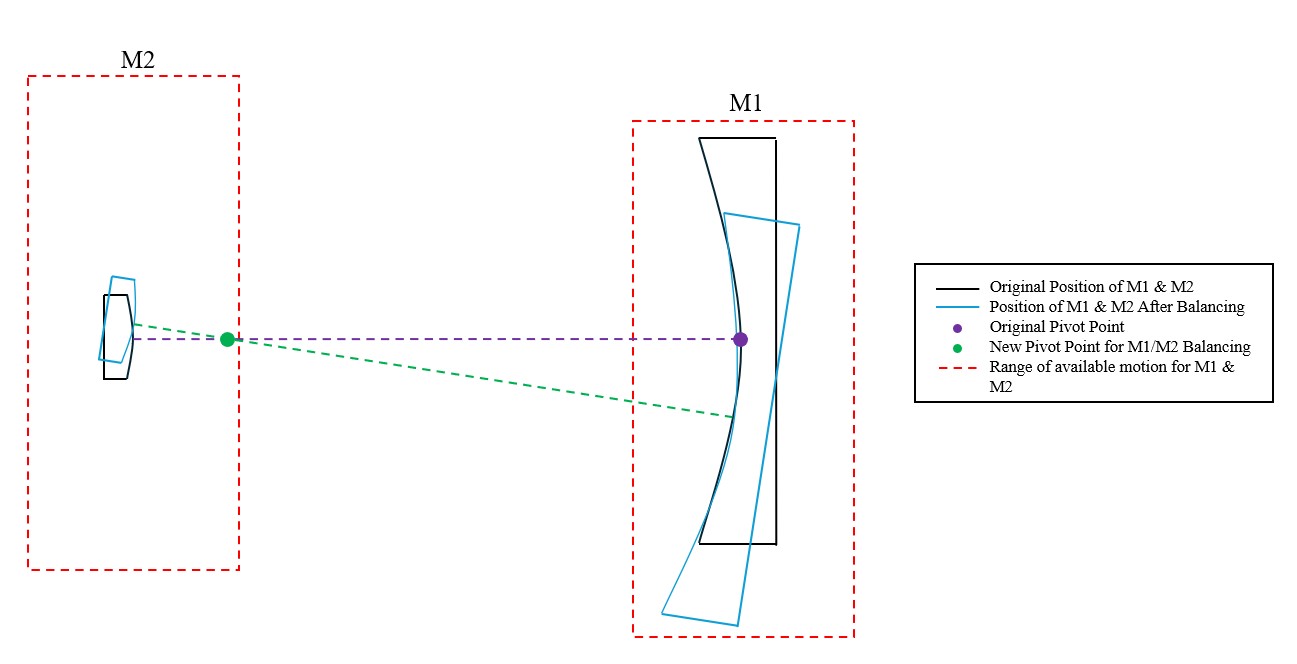}
        \caption{Pivot point orientation between M1 and M2 in the case where M2 has a large range of motion to be able to balance our M1's misalignment.}
    \end{subfigure}
    \caption{Schematic showing how the pivot point position can vary based on how much available range of motion remains on M1 and M2 after the model-based correction step. This schematic shows the pivot point calculation based on one axis but a position is determined for a 3-D coordinate (X, Y, Z) depending on how much motion remains on M1 and M2 after model-based correction.}
    \label{fig:balancing}
\end{figure}

This balancing calculation is done through a geometric process that includes taking the ratio between how much range of motion is remaining on M1 and M2. From that ratio we can determine where a common pivot can be located in relation to the origin point, which is at the vertex of M1. Once a pivot point location is found, the goal of this step is to calculate how much M1 motion is required in relation to the newly defined pivot point to retain the same optical performance relative to M2. A schematic of how the balancing between M1 and M2 will physically translate to the telescope given how much range of motion remains on M1 and M2 along with the misalignment of M2 after model-based correction is shown in Figure \ref{fig:balancing}. Notice how in both schematics, either M1 or M2 approach the limits of their range of motion, hence needing to engage this step of the commissioning alignment.

The result from this step are the newly calculated positions for M1 and M2 that retains optical performance while decreasing the movement on M2. Since the single body of the M1-M2 pair changes the pointing of the telescope, the spot size on the detector plane should be carefully selected and evaluated. Whenever the M2 motion encounters the limit later on in any alignment procedure, we can revisit the balancing step and move M1-M2 based on the updated calculation. In other words, M1's motion is not actively controlled throughout commissioning and will only be moved if balancing is needed.

\subsection{Stochastic Parallel Gradient Descent (SPGD)}

The next and final stage of coarse alignment in the commissioning phase is SPGD. This algorithm was originally developed for model-free optimization in deterministic artificial neural networks and has been used for adaptive phase-distortion correction in white light and active imaging systems \cite{Vorontsov:98}. For our application specifically, we will be using SPGD to drive the finer motion of the five degrees of freedom available on M2 relative to M1 with knowledge from an informed model of the telescope system. Originally, the SPGD algorithm could provide the optimal answer to unknown model problems; however, we utilize the knowledge of a telescope, such as the efficient random perturbation set, to more efficiently survey all available solutions for alignment. The base theory behind the method is to apply random perturbations to a set of control parameters, the magnitude and sign of the perturbations can change each iteration depending on the trend of some defined quality metric \cite{Vorontsov:98}. For this study, the defined control parameters are the five available degrees of freedom on M2 ($[D_x, D_y, D_z, R_x, R_y]$) and the quality metric is the measured average RMS spot size across the detector ($R$) \cite{blomquist2024alignment}. After M1 and M2 are properly balanced, we will only move M2 again at this stage while M1 remains stationary from the position it was left at from balancing.

Since SPGD is the final stage of the commissioning alignment before beginning a finer alignment, the goal of this algorithm is to meet a criterion that is near a 2$\lambda$ peak-to-valley wavefront error where conventional phase retrieval wavefront sensing is available \cite{kev_talk, derby2023integrated}. This criterion roughly translates to 2$\lambda$ peak-to-valley wavefront error or approximately an average RMS spot size of $0.03$ mm across the detector. Although, this criteria is defined for operation, in our studies we also attempt to converge further below this threshold to have guaranteed hand off to the next alignment stage. 

\section{Results}

Following the algorithm development for each individual process and making sure that each stage converges to its alignment goal shown in Figure \ref{commissioning_chart}, we then merged all methods of alignment together to evaluate the robustness against a distribution of random misalignment cases. In practice, the alignment phase that took the most time and would benefit from further optimization is SPGD. 

The main goal of SPGD is to be able to guide the random perturbations of the 5 degrees of freedom on M2 in a way that decreases the RMS spot size in an efficient manner. To do this, we approached troubleshooting from different angles while trying to keep computing requirements low and the logic used simple. Since each SPGD trial produces random perturbation values, after achieving ideal results, it is difficult to recreate the same convergence multiple times. The magnitude of perturbation applied to M2 at each iteration is dictated by a gain parameter. Initial efforts to select gains based on trends or look-up tables has yet to show significant improvements to convergence of the algorithm for multiple misalignment cases. 

All possible misalignment cases are derived from tolerance and error budgets developed specifically for the TMA telescope design used for this study \cite{choi2023approaches}. To test the initial robustness of the algorithm, we use 15 cases from the Monte Carlo simulation. 

\subsection{Commissioning Simulation Results}

\begin{figure}[hbt!]
    \centering
    \includegraphics[width=1\textwidth, height=0.4\textheight]{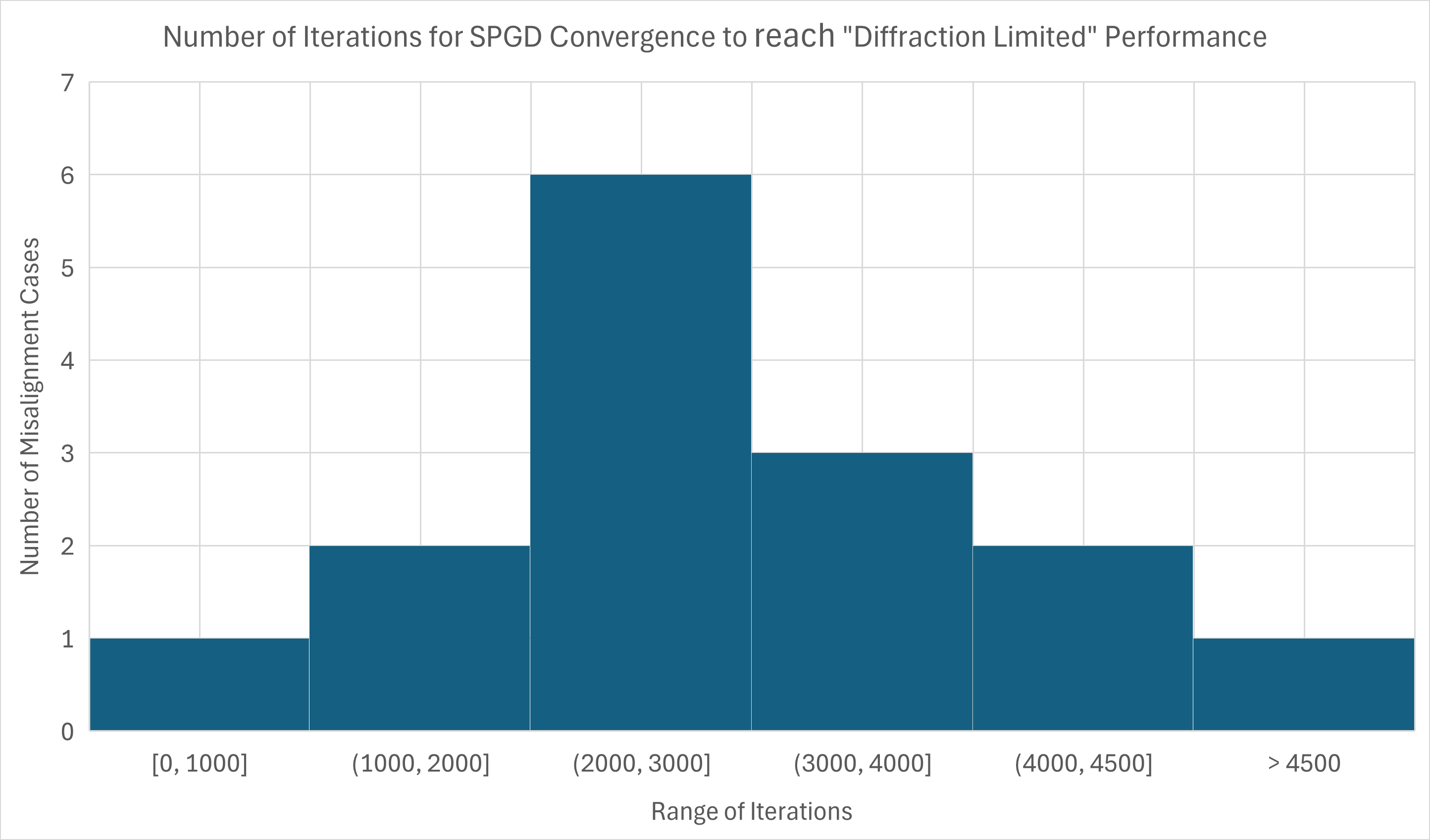}
    \caption{Results showing number of iterations taken to converge to 2$\lambda$ peak-to-valley wavefront error or approximately $0.03$ mm average spot size using SPGD. The number of iterations for convergence is recorded after model-based correction has taken place.}
    \label{spgd_hist}
\end{figure}

When testing the model-based alignment outlined in Section \ref{model_based}, we used a spread of misalignment cases. For testing the entire commissioning algorithm, we will revisit cases taken from the same Monte Carlo simulation and test that the sequence of algorithms, starting from model-based correction, meets the 0.03 mm RMS spot size criteria. The histogram shown in Figure \ref{spgd_hist} highlights the number of iterations taken for the SPGD algorithm to reach convergence after Blind Scanning and model-based correction have taken place.

%  \begin{table}[h!]
%     \setlength{\tabcolsep}{6 pt} % Default value: 6pt
%     \renewcommand{\arraystretch}{1} % Default value: 1
%     \centering
%     \begin{tabular}{|c|c|c|c|} \hline 
%          \textbf{Case Number}& \textbf{Initial RMS Spot Size (mm)} & \textbf{Final RMS Spot Size (mm)} & \textbf{Number of Iterations}\\ \hline 
%          1 & 1.59 & 0.03 & 1494 \\ \hline 
%          2 & 1.90 & 0.03 & 2882 \\ \hline 
%          3 & 2.28 & 0.03 & 3802 \\ \hline 
%          4 & 1.44 & 0.03 & 2979 \\ \hline 
%          5 & 1.88 & 0.03 & 4554 \\ \hline 
%          6 & 1.15 & 0.03 & 3300 \\ \hline 
%          7 & 0.81 & 0.03 & 1883 \\ \hline 
%          8 & 0.86 & 0.03 & 954 \\ \hline 
%          9 & 1.93 & 0.03 & 4438 \\ \hline 
%          10 & 1.80 & 0.03 & 2841 \\ \hline 
%          11 & 1.68 & 0.03 & 2973 \\ \hline 
%          12 & 2.03 & 0.03 & 2873 \\ \hline 
%          13 & 2.41 & 0.03 & 3955 \\ \hline 
%          14 & 1.28 & 0.03 & 2280 \\ \hline 
%          15 & 2.36 & 0.03 & 4134 \\ \hline 
%     \end{tabular}
%     \caption{Results from running commissioning algorithm from model-based correction to SPGD for 15 misalignment cases taken from Monte Carlo simulation in Zemax.}
%     \label{tab:final_spgd}
% \end{table}

\begin{figure}[H]
    \centering
    % Top row: Two subfigures
    \begin{subfigure}{\textwidth}
        \centering
        \includegraphics[width=0.7\textwidth, height = 0.38\textheight]{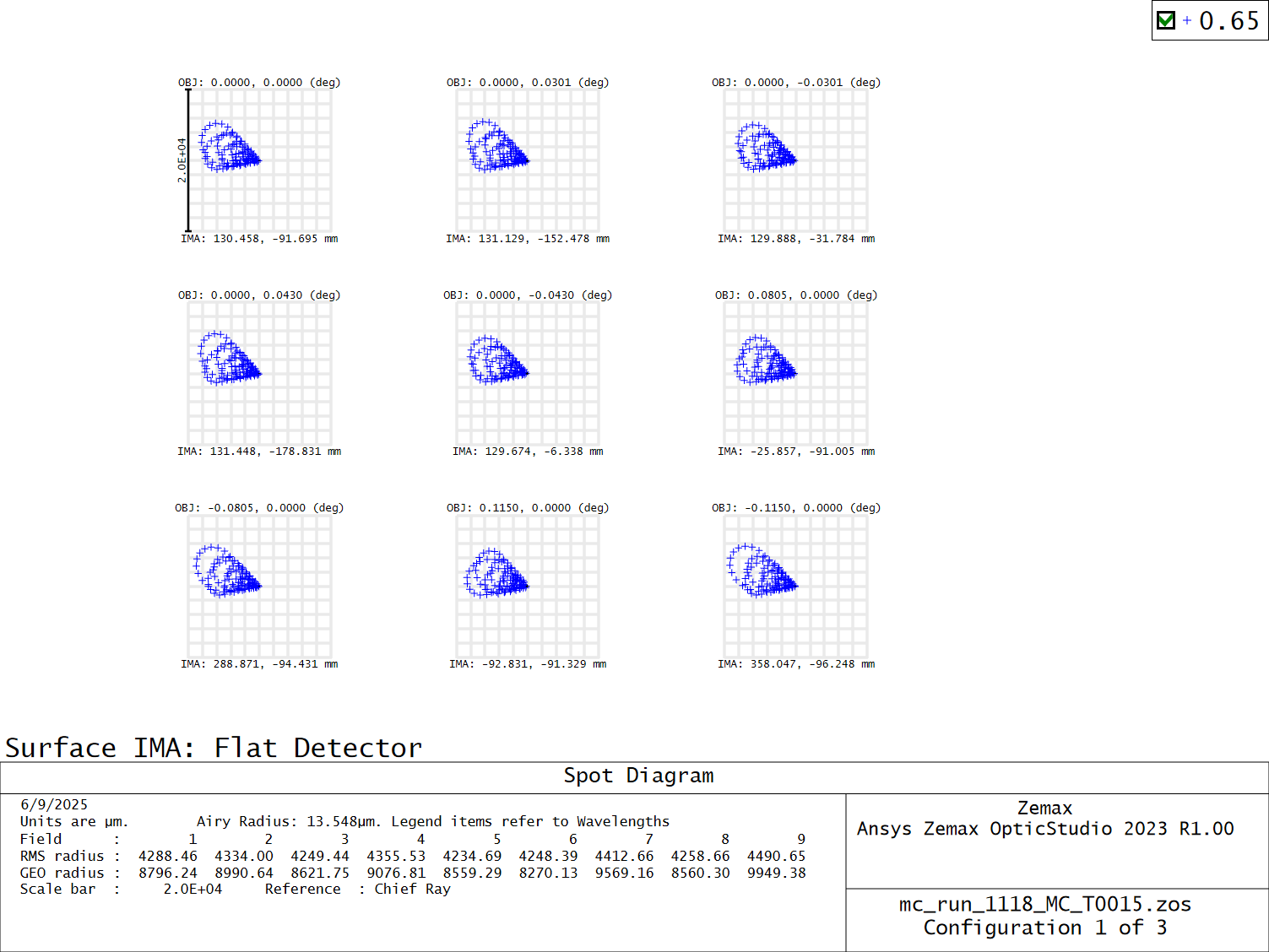}
        \caption{Spot diagrams across all 9 fields before employing commissioning algorithm for correction. The diagrams shown across the 9 field points will roughly correspond to what the detector will see at the stage following Blind Scanning.}
    \end{subfigure}
    \newline
    \begin{subfigure}{\textwidth}
        \centering
        \includegraphics[width=0.7\textwidth, height = 0.38\textheight]{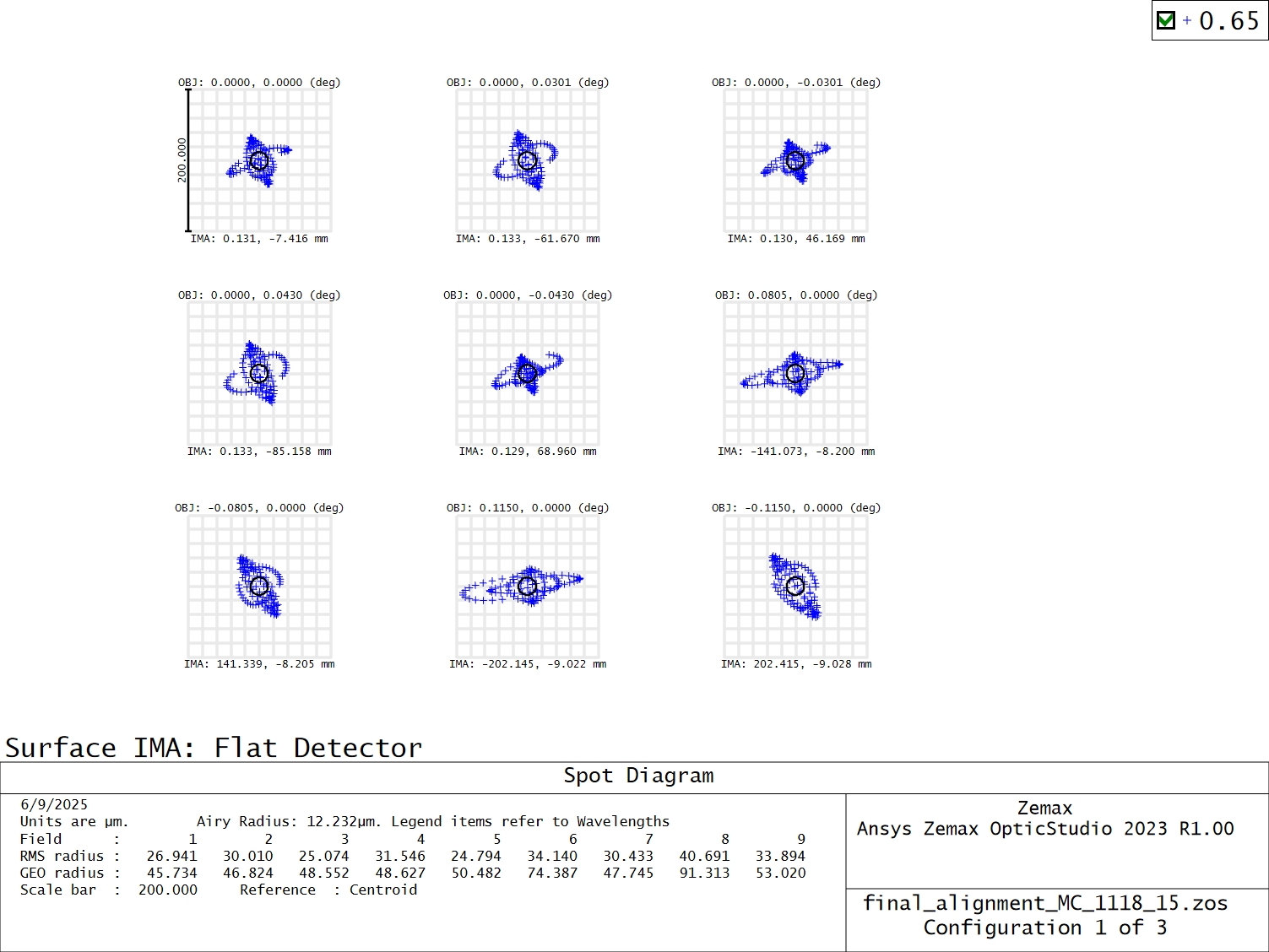}
        \caption{Spot diagrams across all 9 field points after employing commissioning algorithm for correction. These spots roughly correspond to what the detector will see after the telescope system undergoes model-based correction, M1 and M2 balancing (if needed) and finally SPGD.}
    \end{subfigure}

    \caption{Spot diagrams from Zemax before and after commissioning algorithm alignment. Plot (a) shows the spot diagram across 9 field points on the detector before the commissioning algorithm is used to align the telescope. Plot (b) shows the same field points across the detector but after the commissioning algorithm. Notice the decrease in the scaling between the misaligned and aligned spot diagrams by two orders of magnitude, this further confirms the algorithm's ability to decrease spot size. The total field of view spans $\pm$ $0.115^\circ$ in X and $\pm$ $0.043^\circ$ in Y.}
    \label{fig:spot_diagrams}
\end{figure}

Looking at one misalignment case from Figure \ref{spgd_hist}, we ran a spot diagram analysis in Zemax and analyzed the spot diagram before and after the commissioning algorithm. These sets of spot diagrams are shown in Figure \ref{fig:spot_diagrams}.  The black ring for both spot diagrams denotes the Airy disk diameter.  It is not easily seen in the spot diagram before alignment due to high coma present in the system but can be seen in the spot diagram after alignment. 

\newpage

As shown in Figure \ref{spgd_hist}, to reach our criterion before science alignment, 1000 or more iterations are needed of SPGD for multiple misalignment cases. This can conflict with the time the telescope is allotted for actual commissioning time that has yet to be defined. For future work, it will be critical to enhance the convergence rate, so the improved SPGD algorithm can meet the defined criteria in 100 iterations or less ensuring the allotted time for commissioning. Research is already underway to investigate how some methods for accelerating SPGD can be applied effectively to have faster convergence in our system such as Nesterov momentum gradient acceleration and improving how gain values are calculated on each iteration. \cite{nesterov2013gradient}\cite{pauluhn_spectroradiometry_2015}

\section{Future Works and Conclusions}

The results from this study demonstrate that we can use multilayered efficient and robust alignment methods to coarsely align a three-mirror anastigmat telescope. This series of alignment algorithms all solely depend on being able to detect a star on the detector and capture its image. In addition, they have overlapping buffer regions in terms of handover to the next method to build up robustness against an unexpected performance degradation at a certain stage of the alignment. We have shown that this series of algorithms works effectively to converge our system to the current ''diffraction limited'' requirement of the telescope system at commissioning - 2$\lambda$ peak-to-valley wavefront error or roughly $0.03$ mm average RMS spot size across the imaging detector. 

Future works for this study include investigating acceleration methods that can be applicable to SPGD in the context of this telescope and also introducing noise to match reality of system on orbit. At this time, the current TMA design does not have a required time allocation for the commissioning alignment phase but we are aiming to show convergence in 100 iterations or less. We also plan to test this algorithm in a controlled lab environment using an optical system that will mimic the optical performance of the telescope along with testing the proposed alignment scheme using different TMA optical models which will be presented in a future paper.

\section*{ACKNOWLEDGMENTS}

Portions of this research were supported by funding from the Technology Research Initiative Fund (TRIF) of the Arizona Board of Regents and by generous philanthropic donations to the Steward Observatory of the College of Science at the University of Arizona.

\newpage

% References
\bibliography{report} % bibliography data in report.bib
\bibliographystyle{spiebib} % makes bibtex use spiebib.bst

\end{document}